%% file: lrec-coling2024-example.tex
\documentclass[10pt, a4paper]{article}

\usepackage{lrec-coling2024} 
\usepackage{times}
\usepackage{latexsym}
\usepackage{graphicx}
\usepackage{colortbl}
\usepackage{multirow}
\usepackage{amsmath}
\usepackage{amssymb}
\usepackage{pifont}
\title{Enhancing Code Generation Performance of Smaller Models by Distilling the Reasoning Ability of LLMs}

\name{
    \begin{tabular}{c}
    Zhihong Sun$^{1}$, Chen Lyu$^{1*\dagger}$\thanks{$^{*}$Zhi Jin and Chen Lyu are the corresponding authors.}\thanks{$\dagger$This work was done when Chen Lyu was a visiting scholar at Peking University.}, Bolun Li$^{1}$, Yao Wan$^{2}$ \\ Hongyu Zhang$^{3}$, Ge Li$^{4}$, Zhi Jin$^{4*}$ 
    \end{tabular}
}

\address{$^1$School of Information Science and Engineering, Shandong Normal University, China \\
    $^2$Huazhong University of Science and Technology, China \quad $^3$Chongqing University, China \\
    $^4$Key Lab of HCST (PKU), MOE; SCS, Peking University, China \\
         2022021002@stu.sdnu.edu.cn, lvchen@sdnu.edu.cn, libolun118@gmail.com\\
         wanyao@hust.edu.cn, hyzhang@cqu.edu.cn \\
         \{lige, zhijin\}@pku.edu.cn\\}

\abstract{
Large Language Models (LLMs) have recently made significant advances in code generation through the 'Chain-of-Thought' prompting technique. This technique empowers the model to autonomously devise "solution plans" to tackle intricate programming challenges, thereby improving its performance in code generation. Nevertheless, smaller models have been struggling to keep up with LLMs in deducing these plans, adversely affecting their code generation capabilities. Given the considerable size and associated deployment costs, along with concerns about data security, many teams opt for deploying smaller models for code generation. Consequently, there arises a compelling need for transferring LLMs' code generation reasoning abilities to the smaller models. In this paper, we propose the CodePLAN framework, which aims to transfer LLMs' reasoning capabilities to smaller models through distillation. We adopt a multi-task learning approach, jointly undertaking code generation and solution plan generation tasks, to enhance the code generation capabilities of the smaller model. 
To ensure the superior quality of the solution plans, we advocate for the utilization of backward reasoning and plan sampling strategies. 
Our experiments show that in comparison to the conventional fine-tuning approach, our approach improves the smaller model's code generation performance (measured in pass@1 metric) by over 130\% on the challenging APPS benchmark.
}

\begin{document}

\maketitleabstract

\input{sections/01_instroduction}
\input{sections/02_related}
\input{sections/03_CodePLAN}
\input{sections/04_experiments}
\input{sections/05_discussion}
\input{sections/06_conclusion}
\input{sections/07_limitations}
\input{sections/08_acknowledgments}

\section{Bibliographical References}\label{sec:reference}

\bibliographystyle{lrec-coling2024-natbib}
\bibliography{lrec-coling2024-example}

\appendix
\clearpage
\input{sections/09_appendice}


\end{document}

%% file: sections/01_instroduction.tex
\section{Introduction}

Automatic code generation has a history spanning decades, aiming to create executable programs from problem specifications~\citep{backus1957fortran,waldinger1969prow,manna1971toward}. As artificial intelligence technology rapidly advances, the application of neural network techniques in intelligent code generation is increasingly gaining attention in the field of software engineering~\citep{ling2016latent, yin2018tranx, lyu2021embedding}. Recently, large language models (LLMs) such as ChatGPT~\citep{ChatGPT} have made significant advances in code generation owing to their superior reasoning capabilities. 
However, deploying these mammoth models comes with significant computational, time, and financial demands, coupled with data and security risks. Consequently, many enterprises and teams still prefer more manageable, smaller models.

\begin{figure}
    \centering
    \includegraphics[width=0.95\linewidth]{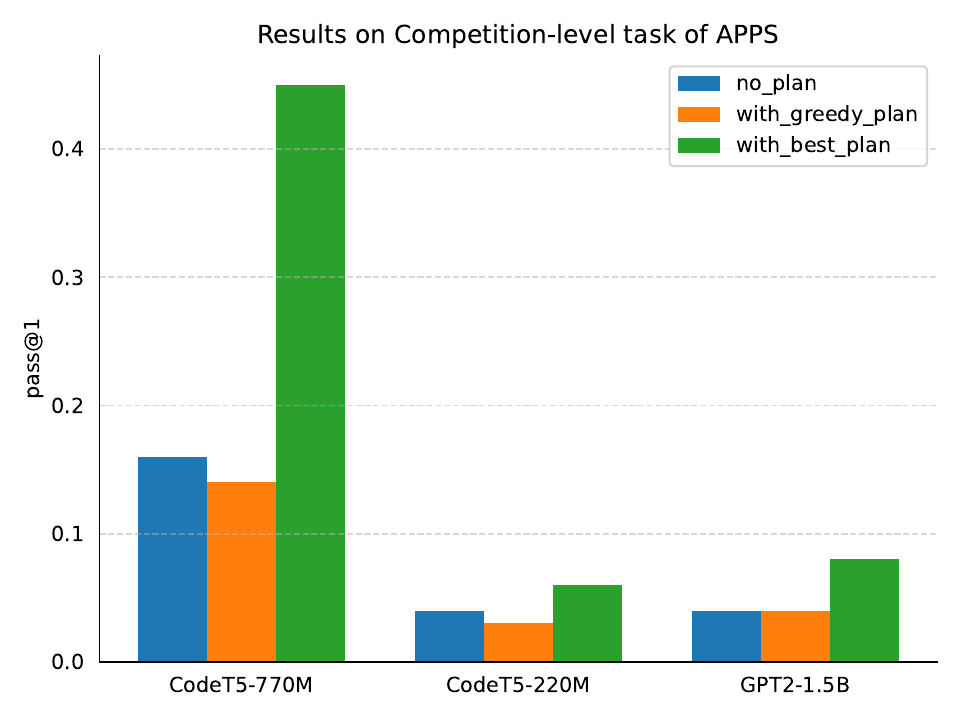}
    \vspace{-3mm}
    \caption{Comparison results of different models without solution plans, spliced LLM greedy generated solution plans and the best quality solution plans as prompt, where all models were fine-tuned on the APPS train dataset.}
    \label{fig: figure1}
    \vspace{-4mm}
\end{figure}

In the realm of code generation, smaller models lag in reasoning capabilities compared to LLMs, leading to challenges with complex programming tasks. Our empirical studies highlight the exceptional in-context learning (ICL) of LLMs. By utilizing "Chain-of-Thought (CoT)"~\citep{wei2022chain} as human-defined solution steps, LLMs can bolster their reasoning, allowing them to craft solution plans from these in-context examples. This methodology elevates LLMs' problem-solving accuracy and is notably effective in code generation~\citep{jiang2023self,huang2024knowledge}. However, while CoT strategies shine with massive-parameter models, smaller models, even after fine-tuning, struggle in deriving CoT-based solution plans due to ICL and reasoning constraints. Yet, we have observed that a smaller model, around 1B parameters in size, when fine-tuned and given both problem description and a precise CoT-based solution plan (labeled as "best plan"), sees a substantial boost in code generation capabilities, as shown in Figure \ref{fig: figure1}.

This finding prompted us to further explore strategies for providing smaller models with a ``best plan'' when addressing programming tasks. However, we face two significant challenges:
\textbf{1) Dependency on Large Language Models (LLMs) during inference.} Utilizing LLMs to generate solution plans for smaller models might be pragmatic, but it becomes impractical if the smaller model consistently relies on LLMs during inference.
\textbf{2) Securing accurate, high-quality solution plans.} Solution plans are primarily procured either manually by experts or automatically by LLMs. While expert-curated plans are typically more reliable, their high costs make them less feasible for automated code generation. In contrast, most LLM-generated plans, directly derived from problem descriptions, often do not meet the desired quality. Notably, even state-of-the-art LLMs like ChatGPT can produce plans that are not consistently accurate, leading not only to missed enhancements but potential performance regressions in smaller models, as depicted in Figure \ref{fig: figure1}.



To address these challenges, we introduce ``CodePLAN'', a novel multi-task plan-based framework designed to enhance the code generation for smaller models by distilling LLMs' reasoning ability. Essentially, CodePLAN utilizes multi-task learning to imbue smaller models with LLMs' reasoning capabilities, allowing them to autonomously develop solution plans and generate code. Central to CodePLAN's effectiveness is the precision of these solution plans, both in training and inference. Thus, we innovate two techniques: "back reasoning" and "plan sampling", which respectively enhance the quality of plans during LLM distillation and during CodePLAN's own inference.


Specifically, to tackle the first challenge, we conceptualize LLMs as "teachers" and smaller models as "students", with the objective of distilling the teacher's reasoning capabilities into the student. We employ a multi-task training framework, using solution plans from LLMs and actual codes as supervisory signals. This framework emphasizes two tasks: 1) Code generation, which develops the smaller model's coding skills, and 2) Plan generation, aiming to distill LLMs' reasoning prowess. Leveraging this strategy, the smaller model's code generation performance improves notably. While it leans on LLMs during training, it operates autonomously during inference. At this stage, capitalizing on its refined skill to generate solution plans, the model uses its plans to enhance the code generation process, optimizing its output potential.
For the second challenge, we guide LLMs to create solution plans based on actual codes using a "back reasoning" approach. This deviates from methods by ~\citet{jiang2023self} and ~\citet{li2023think}, who rely solely on problem descriptions. Our method prioritizes obtaining top-tier solution plans (refer to section \ref{sec:3.1}), a claim supported by our empirical data (see Table \ref{tab: Tabel 7}). 
Nonetheless, smaller models still confront a similar obstacle during the inference phase - the inability to directly generate correct solution plans. To address this problem, we take inspiration from the process that programmers use to solve complex programming problems. The process of programmers in solving complex programming problems is actually a process of continuous trial and error of thinking, where the correctness of thinking has been verified by writing code according to the constructed thinking until the problem is solved.  Consequently, in the inference phase, we introduce a technique called "plan sampling" to simulate a programmer's problem-solving. The key was to ensure efficiency while targeting quality solutions. To this end, we craft a strategy using limited sampling and concise unit tests for each sampled solution plan. This makes "plan sampling" both lightweight and highly effective.


We executed a comprehensive series of experiments on two distinct streamed code generation datasets, namely APPS and MBPP. Our novel approach, in comparison to standard finetune methods, considerably enhances the code generation proficiency of the model, most notably improving the pass@1 metric on the APPS dataset by over 130\%. To the best of our knowledge, this study is the first exploration of distilling the reasoning ability of LLMs to improve code generation in smaller models. 
Our codebase is publicly accessible at: \url{https://github.com/sssszh/CodePLAN}.

%% file: sections/02_related.tex
\section{Related Work}
\paragraph{Code Generation.} With the advent of transformer~\citep{vaswani2017attention} and the development of pre-training techniques~\citep{devlin2018bert}, more and more pre-training models are applied in the field of code generation. For instance, open-source code models like CodeT5~\citep{wang2021codet5}, CodeT5+~\citep{wang2023codet5plus}, CodeGen~\citep{nijkamp2022codegen}, PolyCoder~\citep{xu2022systematic}, InCoder~\citep{fried2022incoder}, StarCoder~\citep{li2023starcoder}, as well as general-purpose language models such as GPT-J~\citep{gpt-j}, GPT-Neo~\citep{black2021gpt} have demonstrated substantial performance in code generation tasks. 

The dominant approaches in code generation mainly involve fine-tuning pre-trained code generation models using supervised learning~\citep{APPS} or reinforcement learning (RL)~\citep{li2022competition, le2022coderl, shojaee2023executionbased, li2024ircoco}. However, neither supervised nor reinforcement learning fine-tuning allows the model to learn reasoning well. Moreover, RL-based approaches decompose code generation into sequences of token-generating actions, which may limit the model to learn reasoning ability due to the lack of high-level thinking. Different from these methods, we achieve high-level thinking in smaller models by distilling the reasoning abilities of LLMs into them.
\paragraph{Chain-of-Thought (CoT).}With the advent of large language models, such as ChatGPT ~\citep{ChatGPT} and GPT4~\citep{GPT-4}, and the evolution of CoT prompting techniques~\citep{wei2022chain, wang2022self}, an increasing number of researchers have committed themselves to identify strategies that effectively augment the emergent capabilities of LLMs~\citep{shum2023automatic, zhou2022least}.~\citet{jiang2023self} proposed a ``self-plan'' approach, leveraging the inherent reasoning abilities of LLMs to sequentially decompose and solve problems. This methodology has yielded promising results for fundamental programming tasks. However, these CoT prompt-based techniques are predominantly applicable to models with many parameters (e.g., 100 B or more) and are less suitable for models with fewer parameters, which lack inferential solid interpretation abilities to decompose complex problems independently.~\citet{ho2022large} and ~\citet{hsieh2023distilling} employed a novel strategy of using inference interpretations generated by LLMs as supervised signals to train smaller models, with the aim of enhancing their performance on simple natural language processing (NLP) tasks.

In summary, the ``self-plan'' approach proposed by~\citet{jiang2023self} relies heavily on the inherent reasoning ability of LLMs. Different from methods that stimulate the inherent reasoning abilities of LLMs themselves, our methodology utilizes solution plans generated by LLMs as supervised signals for training smaller models, distilling the reasoning ability of LLMs into small models, it can reduce the expensive cost of deploying LLMs. Previous studies~\citep{ho2022large,hsieh2023distilling} have leveraged the reasoned interpretation of LLMs to enhance the performance of smaller models on simple NLP tasks. However, unlike these simple NLP tasks, code generation is a much more complex task, and the difficulty of obtaining high-quality solution plans prevents these methods from being directly applied to the field of code generation. 
\begin{figure}
    \centering
    \includegraphics[width=0.55\linewidth]{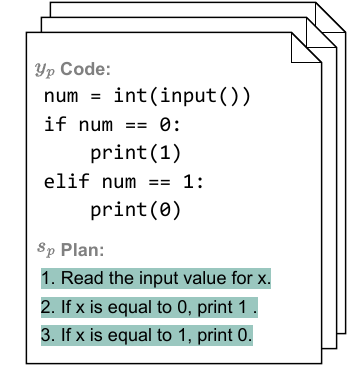}
    \vspace{-3mm}
    \caption{We use the prompt to allow LLM to reason backwards a solution plan from the code written by the programmer (highlighted in green).}
    \label{fig:CoT-example}
    \vspace{-5mm}
\end{figure}

%% file: sections/03_CodePLAN.tex
\section{CodePLAN}
\begin{figure*}
    \centering
    \includegraphics[width=0.85\linewidth]{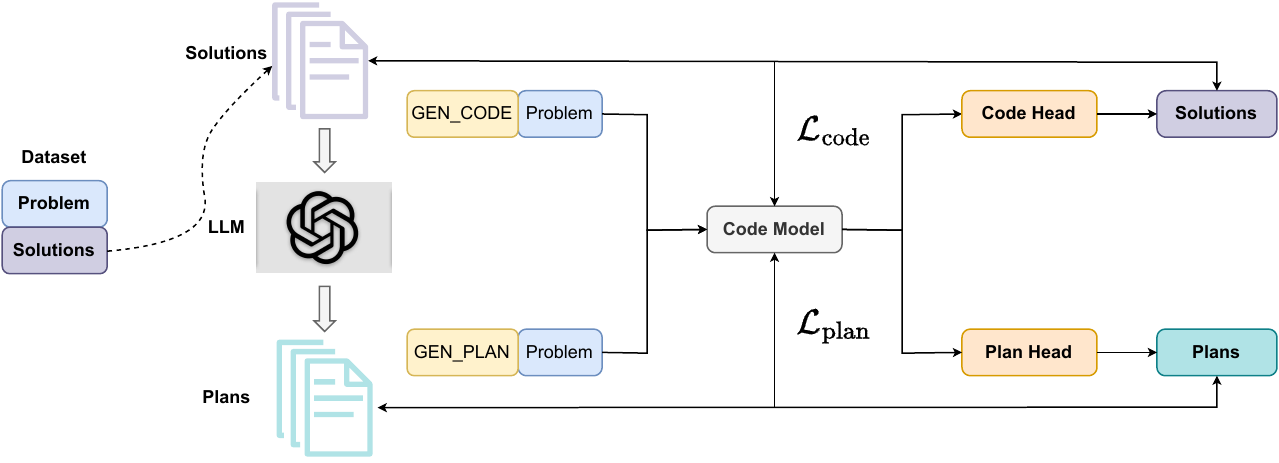}
    \caption{\textbf{Our framework for the training phase of CodePLAN:} backward reasoning from solutions via LLM about the programmer's solution plan at the time of solving this programming problem, and using these solution plans and solutions to fine-tune the code generation model in an alternating multi-task fashion.}
    \label{fig:optimization phase}
\end{figure*}
In this section, we provide a comprehensive exposition of the core principles underlying CodePlan. First, we outline how CodePlan extracts distilled knowledge from LLMs, with a specific emphasis on the key ingredient termed "solution plans". Following this, we demonstrate the detailed training process by which CodePlan leverages these "solution plans" within a multi-task learning framework. Lastly, we describe in-depth how CodePlan, during its inference phase, utilizes its self-generated "solution plans" to facilitate code generation.
\subsection{Generating Plans through LLM}
\label{sec:3.1}
Contemporary research reveals that LLMs have the capacity to generate high-quality inference steps for certain rudimentary NLP tasks, thereby interpreting the solutions it produces~\cite {wei2022chain}. However, within the domain of code generation, LLM does not guarantee the generation of high-quality inference steps for intricate programming challenges. This necessitates the exploration of an effective methodology to uncover these high-quality solution plans.

Most LLM-based plan generation methods use a "forward reasoning" strategy, leveraging CoT examples to deduce plans from problem descriptions. However, our empirical studies for code generation tasks suggest that "backward reasoning" — deducing from the given solution/code — often produces higher-quality plans. 
To facilitate this, we establish a dataset $D$, comprised of paired elements $(x_{i}, y_{i})$, where $x_{i}$ represents the problem description, and $y_{i}$ represents the solution to $x_{i}$. For complex programming tasks, the LLM may not be able to generate high-quality solution plans directly from $x_{i}$. In contrast, as $y_{i}$ represents the solution to $x_{i}$ as authored by the programmer, it intrinsically encompasses the programmer's solution plan for addressing the problem. Consequently, we infer the solution plan $s_{i}$, written by the programmer when composing $y_{i}$ to solve $x_{i}$, by reasoning backwards from $y_{i}$. The solution plan we deduce backwards from $y_{i}$ is typically superior to the solution plan obtained directly from $x_{i}$ using the LLM~(see section~\ref{sec: 4.5}). We guide the LLM to generate solution plans based on provided prompt $(y_p,s_p)$, utilizing the prompt template shown in Figure~\ref{fig:CoT-example}. For a new $y_{i}\in D$, the LLM emulates the prompt $(y_{p},s_{p})$ to reason backwards a solution plan $s_{i}$ for $y_{i}$.

\subsection{Training Model with Plans}
We initially outline the methodology for training the base model using the solution plans. In this procedure, we employ the intermediate solution plans, generated by LLMs, as a novel fine-tuning task assigned to the base model. The specifics of this training process are graphically depicted in Figure~\ref{fig:optimization phase}.

In conventional fine-tuning strategies, the base code generation model typically aims to minimize the cross-entropy loss between the generated code and the target code, serving as the primary training objective:
\begin{equation}\mathcal{L}_{\text {code }}\left(\theta_1\right)=-\sum_t \log p_{\theta_1}(w_t \mid w_{1: t-1}, D)\end{equation}
where $D$ represents the problem description and $W=(w_1,...w_t)$ represents the ground truth code.

Nonetheless, this conventional fine-tuning strategy fails to equip the model with inferential proficiency. To endow smaller models with the LLM's capability to decompose intricate problems, we add a training task -distilling reasoning ability from LLM - generating solution plans. This task is executed to minimize the cross-entropy loss between the solution plans generated by the model and those generated by the LLM:
\begin{equation}\mathcal{L}_{\text {plan }}\left(\theta_2\right)=-\sum_t \log p_{\theta_2}(s_t \mid s_{1: t-1}, D)\end{equation}
where $D$ represents the problem description and $S=(s_1,...s_t)$ represents the solution plan generated by LLM.

This approach not only equips the model with code generation capabilities but also enables it to generate intermediate solution plans. Within this training workflow, we utilize an alternating training strategy to fine-tune our model, distinguishing between the two tasks using two unique characters: $[GEN\_CODE]$ and $[GEN\_PLAN]$. Given the stark differences between program language and natural language, we modified our model by incorporating a new ``plan head'' at the end of the base model to generate solution plans. The total loss function optimized in our model is:
\begin{equation}
\mathcal{L}=(1-\lambda)\mathcal{L}_{\text {code }}+\lambda \mathcal{L}_{\text {plan }}
\end{equation}
where $\lambda$ is a hyperparameter that regulates the weight assignment for the loss of the two tasks. In our experimental setup, $\lambda$=0.5.
\begin{figure*}
    \centering
    \includegraphics[width=0.90\linewidth]{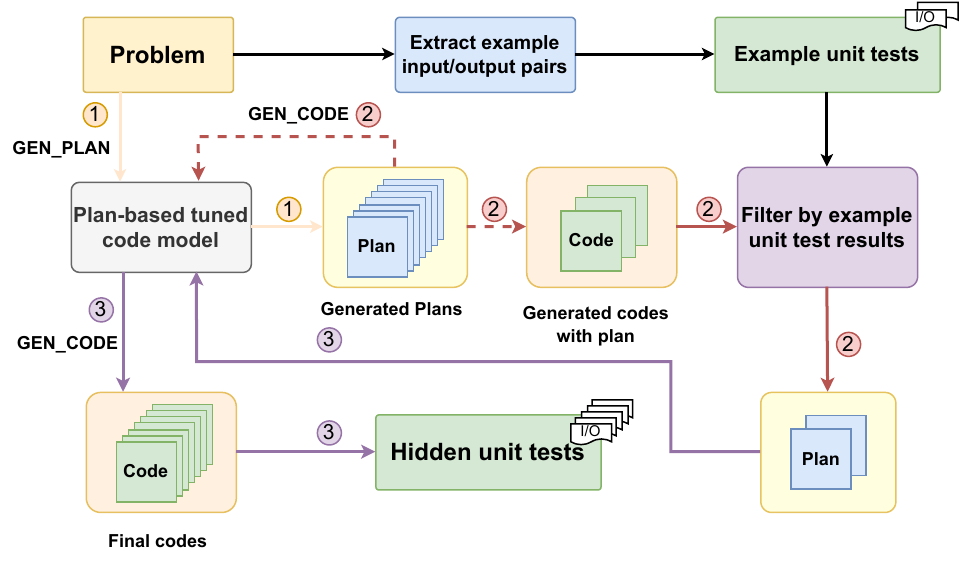}
    \caption{\textbf{The inference phase schematic comprises three stages: }\ding{172} Initially, the model formulates candidate solution plans based on the provided problem description. \ding{173} Subsequently, as indicated by the dashed line, solution plans generated in Stage \ding{172} are integrated with the problem description for code generation.  Candidate solution plans are chosen based on the evaluation outcomes of the code generated through example unit tests. \ding{174} Ultimately, the selected high-quality solution plan is used as a prompt, integrated within the problem description for a new cycle of code generation.}
    \label{fig:inference phase}
    \vspace{-1mm}
\end{figure*}
\subsection{Inferencing with Plans}
Leveraging a multi-task fine-tuning approach, our framework enables the base model to generate both code and solution plans. In this context, we detail how, during the inference phase, the solution plans produced by the model enhance code generation, as illustrated in Figure~\ref{fig:inference phase}.
\paragraph{Plan Sampling.}
As shown in Figure~\ref{fig:inference phase}, the inference phase of CodePLAN is delineated into three specific stages. Initiated in the first stage, we use the fine-tuned model to generate the solution plans. The input of the model consists of the $[GEN\_PLAN]$ label and the problem description $x_{i}$, and its output is the solution plan $s_{i}$. However, the utilization of a greedy decoding strategy is insufficient to assure the precision of the solution plans. This deficiency prompted us to consider the methods by which programmers tackle complex competition problems: in such scenarios, a plethora of potential solution plans are conceived, with code being written and subsequently verified through unit tests. 

Accordingly, we incorporated a novel strategy - ``plan sampling'' - to get the correct solution plans. This approach permits the sampling of multiple solution plans per problem, thereby encouraging the model to ideate akin to a programmer while acknowledging that complex programming problems may have multiple solutions. For the second stage, the model utilizes the $[GEN\_CODE]$ label, the problem description $x_{i}$, and the solution plan $s_{i}$ as input, and then proceeds to generate a small volume of codes $y_{i}=\{y_{i_{1}}, y_{i_{2}},...,y_{i_{n}}\}$ in accordance with the solution plan, where $n$ is limited to 10. This process can be formally defined as: $y_{i}\longleftarrow f(x_{i},s_{i})$, where $f$ represents the base model. 
Considering time constraints, we sample up to 20 solution plans. The quality of a solution plan is indicated by the number of generated codes that successfully pass the example unit tests $t_{i}$ — typically, only one or two as specified in the problem description.
This evaluation metric is formalized as $score(s_i) = \sum_{y_{i_{n}} \in y_i} \delta(y_{i_{n}}, s_i, t_i)$. Here, $\delta(y_{i_{n}}, s_i, t_i)$ represents whether the code $y_{i_{n}}$, guided by the plan $s_i$, passes the example unit tests $t_i$, defined as: $\delta(y_{i_{n}}, s_i, t_i):= \begin{cases}1, & \text { if } y_{i_n} \text { passes} \operatorname t_i \\ 0, & \text { otherwise }\end{cases}$.
Consequently, the solution plan correlating with the highest number of successful code tests is deemed as the highest quality, formally captured by $s_{i}=argmax_{s_i \in S}(Score(s_{i}))$. Transitioning to the third stage, our framework elevates the model's code generation capability using the chosen top-tier solution plan. Codes generated within this enhanced framework are further assessed by hidden unit tests, which are more rigorous than the example unit tests, often capturing edge cases or extreme instances of the code's functionality.


%% file: sections/04_experiments.tex
\section{Experiments}

\input{tabels/01_table_1}
\subsection{Experiment Setup}
\paragraph{Dataset and Models.}In this study, we evaluate our approach on two mainstream code generation datasets: (1) \textbf{APPS}~\citep{APPS}.  (The dataset was collected from several programming competition platforms (e.g. Codeforces, LeteCode, etc.) with 10,000 problems, of which 5,000/5,000 problems were divided for training/testing and divided into three levels according to the difficulty of the problems, introductory level, interview level, and competition level. (2) \textbf{MBPP}~\citep{austin2021program}. The dataset consists of 974 programming problems constructed from crowdsourcing, with 374/90/500 problems divided for training/validation/testing and 10 reserved for few-shot prompt learning. We choose two of the most popular code generation models, CodeT5-770M~\citep{wang2021codet5} and CodeGen-350M~\citep{nijkamp2022codegen}, to validate the effectiveness of our approach. And the solution plans we use for training are from OpenAI's GPT-3.5-Turbo API~\citep{ChatGPT}. 
\paragraph{Metric.}To evaluate the functional correctness of generated codes, we followed the previous works~\citep{APPS, chen2021evaluating} using pass@k as the evaluation metric. This metric measures the functional correctness of the code by executing unit test cases. For each problem sampled to generate n>=k copies of code, the number of correct codes c<=n, pass@k metric is calculated as follows:
\begin{equation}
\operatorname{pass} @ k=\underset{\text { Problems }}{\mathbb{E}}\left[1-\frac{\left(\begin{array}{c}
n-c \\
k
\end{array}\right)}{\left(\begin{array}{l}
n \\
k
\end{array}\right)}\right]
\end{equation}
In our experimental setup, we sample 100 copies of the code for each problem to compute pass@$\{1, 5, 100\}$

\paragraph{Training/Inference Setting.} For the training phase associated with the APPS dataset, we adhered to the data preprocessing structure as delineated in the original paper~\citep{APPS}. The established maximum lengths for the source sequence and the target sequence were 600 and 512, respectively. The batch size was configured to 32, and the learning rate was specified at 2e-5, a learning rate decay of 0.05. The fine-tuning process was executed 10 epochs. When approaching the MBPP dataset, we remained consistent with the data preprocessing methodology laid out in the original paper~\citep{austin2021program}. The respective maximum lengths for the source sequence and the target sequence were set at 350 and 300. Both the batch size and the learning rate mirrored the parameters established for the APPS dataset. Importantly, we implemented a total of 50 rounds of fine-tuning for the MBPP, to account for the more limited number of training sets within this dataset. In the inference stage, we employed temperature sampling for both APPS and MBPP, with respective temperature settings of 0.6 and 1.2. For each problem, we stipulated the generation of 100 instances of the code and 20 solution plans.
\subsection{Experimental Results on APPS}

We evaluated our models and compared them with several baseline models, which include GPT-2~\citep{radford2019language}, GPT-Neo~\citep{black2021gpt}, GPT-3~\citep{brown2020language}, CodeX~\citep{chen2021evaluating}, CodeT5~\citep{wang2021codet5}, CodeT5+~\citep{wang2023codet5plus}, StarCoder~\citep{li2023starcoder} and CodeGen~\citep{nijkamp2022codegen}. Note that all models except CodeX and GPT-3 are fine-tuned on APPS.
As illustrated in Table \ref{tab:tabel1}, CodePLAN notably bolsters the code generation competency of the model, surpassing models equipped with several folds the number of parameters. Specifically, across all levels of the APPS benchmark, CodePLAN secures an impressive gain of over 130\% in the pass@1, as compared to the standard fine-tuning process. Moreover, our method manifests considerable improvements in the pass@5 and pass@100. It's worth emphasizing that the enhancement engendered by our method on the pass@1 metric is significantly more pronounced than on the pass@100. This denotes that CodePLAN significantly escalates the likelihood of the model generating correct code for the identical question. Furthermore, CodePLAN can generate a larger volume of correct codes than alternative methods, thereby proving advantageous for subsequent post-processing tasks like code ranking.  Interestingly, the relative improvement of CodePLAN on complex, competition-level questions surpasses that on introductory-level and interview-level questions, indicating that CodePLAN empowers smaller models to solve complex programming problems with reasoning capabilities.
\input{tabels/03_table_3}
\input{tabels/04_table_4}

\input{tabels/05_table_5}
\subsection{Comparative Analysis of Various Training Approaches}

In this section, we compare various training techniques on APPS and MBPP, namely standard fine-tuning, CoT fine-tuning, RL-based fine-tuning, and CodePLAN without Plan Sampling (abbreviated as CodePLAN w/o PS). Across these methods, a consistent base model is employed. CoT fine-tuning, inspired by existing research ~\citep{ho2022large}, is not typically used for code generation. For this method, we merge the solution plan with the code to create a target sequence. During inference, the model produces a "CoT + Code" output, with the Code segment extracted for assessment. For the RL-based fine-tuning, our reference point is the CodeT5 checkpoint released by CodeRL~\citep{le2022coderl}, a framework that harnesses RL training for code generation.

\textbf{Result on APPS}. On the APPS benchmark, we conducted this experiment using CodeT5 770M~\citep{wang2021codet5} and CodeGen 350M~\citep{nijkamp2022codegen} as base models. Table~\ref{tab: tabel3} presents the comparative results of CodePLAN w/o PS alongside various fine-tuning methodologies on the APPS benchmark. We can find that the code generation ability of smaller base models is improved by distilling the reasoning ability of the LLM, and that this approach outperforms other fine-tuning methods on all difficulty levels of the APPS benchmark. 
Compared to standard fine-tuning and RL-based fine-tuning, the inference ability of the smaller model is improved by distilling the inference ability of the LLM thus indirectly improving the code generation ability of the base model. In contrast, standard fine-tuning and RL-based fine-tuning methods lack high-level thinking as a supervisory signal and are not beneficial for improving the reasoning ability of smaller models.
While CoT fine-tuning has proven its mettle in simpler NLP tasks~\citep{ho2022large}, our experiments reveal its direct application to the intricate realm of code generation to be less impactful. In this context, CodePLAN demonstrates a significant edge.

\textbf{Result on MBPP}. We also conducted this experiment on MBPP, where we followed the experimental setup of the original paper~\citep{austin2021program} and also used the CodeT5-770M and CodeGen-350M as the base models. The outcomes are presented in Table \ref{tab: Tabel 4}. Mirroring findings from the APPS benchmark, CodePLAN (w/o PS) consistently outperforms both standard fine-tuning and CoT fine-tuning methods by leveraging the distilled reasoning capabilities of the LLM.

\subsection{Impact Analysis of Varying Solution Plan Sample Sizes}
Table~\ref{tab: Tabel 5} presents the results of ablation experiments, examining the effect of varying the number of sampled solution plans during inference. In this setup, the model-generated solution plan  $s_{i}$  is appended to the problem description $x_{i}$ to guide the code generation process $y_{i}\longleftarrow f(x_{i},s_{i})$.  Interestingly, with N=1, where the base model generates a single solution plan using greedy decoding, the code generation performance doesn't improve. It might even degrade compared to when no solution plan is used (N=0, represented by $y_{i}\longleftarrow f(x_{i})$). This indicates that solution plans derived via greedy decoding may lack precision. However, employing multiple samplings to select quality solution plans can significantly enhance the model's code generation efficacy. Moreover, a greater sampling quantity increases the likelihood of identifying a more accurate solution plan.
\begin{figure}
    \centering
    \includegraphics[width=0.98\linewidth]{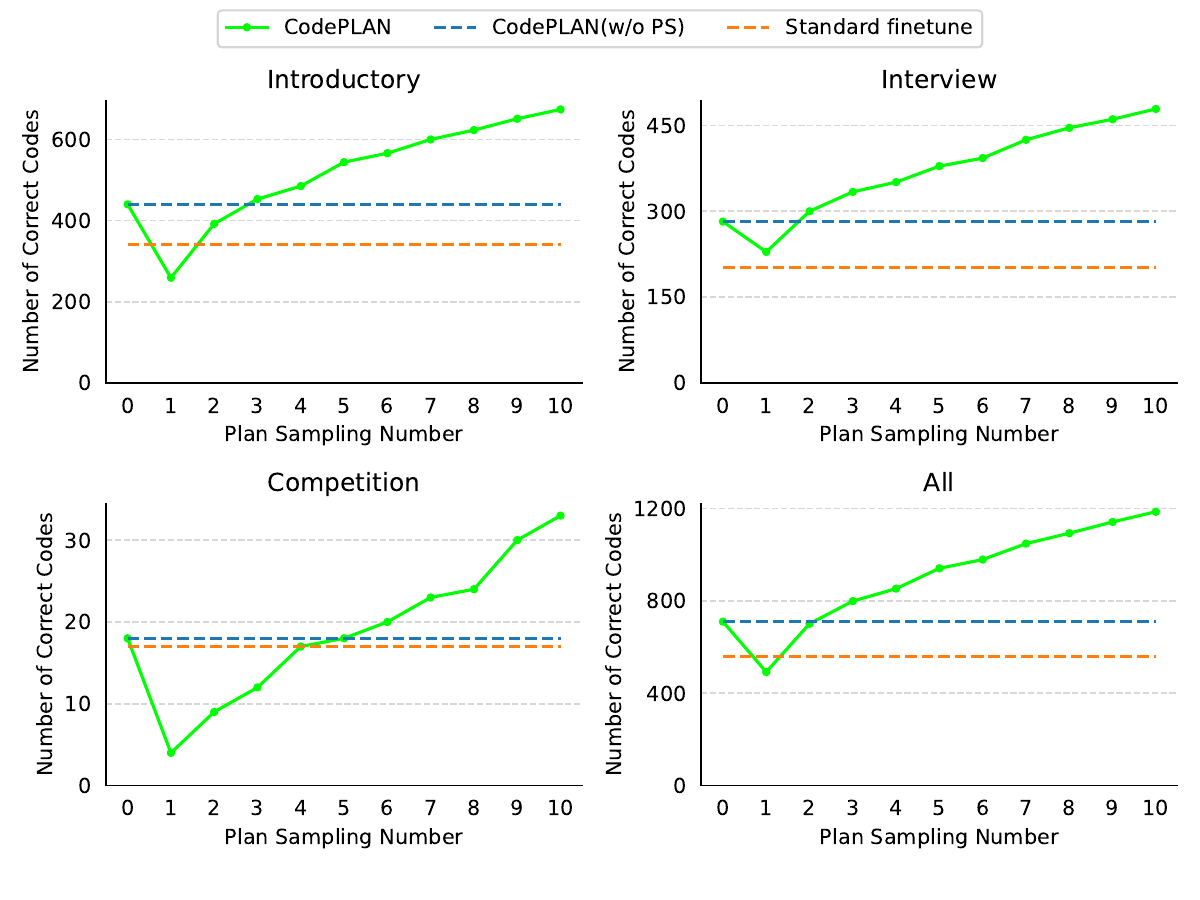}
    \vspace{-3mm}
    \caption{Results of different number of solution plans on the number of correct codes generated.}
    \label{fig:Number plans}
    \vspace{-4mm}
\end{figure}
Figure~\ref{fig:Number plans} depicts how sampling solution plans of varying quantities and difficulties impacts the count of accurately generated codes, evaluated against the APPS dataset. Across all difficulty levels, it is evident that, with N=2, there's an uptick in correct codes relative to the "Standard Finetune". By N=3, the performance even eclipses that of "CodePLAN(w/o PS)". These findings underscore that a minimal sampling of solution plans can effectively yield high-quality selections. This confirms that our approach not only amplifies the code generation prowess of smaller models but also refines their inference accuracy.
\subsection{Evaluation of LLM-Generated Training Data Quality}
\label{sec: 4.5}
In this subsection, we delve into the quality analysis of training data formulated by the LLM. Assessing the quality of solution plans directly generated by the LLM poses challenges. Instead, we resort to an indirect method, evaluating the quality of codes generated under the guidance of these solution plans. For this assessment, we utilize CodeT5 770M, which underwent standard fine-tuning on the APPS dataset.  The results, presented in Table ~\ref{tab: Tabel 7}, compare the quality of LLM-generated solution plans from problem descriptions ("Problem to Plan") and those derived from backward reasoning using ground truth codes ("Code to Plan"). The label "No-Plan" indicates scenarios where LLM-generated plans were not used as auxiliary guidance. Our findings reveal that solution plans derived from backward reasoning using ground truth codes surpass in quality those generated directly from intricate problem descriptions. This likewise indirectly ensures the quality of our distilled data.
\input{tabels/07_table_7}

%% file: tabels/01_table_1.tex
\begin{table*}[t]
\centering
\small

\resizebox{1.0\textwidth}{!} {
\begin{tabular}{lc|cccc|cccc|cccc}
\hline
\multicolumn{1}{c}{}                        &                        & \multicolumn{4}{c|}{\emph{Pass@1}}                                       & \multicolumn{4}{c|}{\emph{Pass@5}}                                        & \multicolumn{4}{c}{\emph{Pass@100}}                                        \\ 
\cline{3-14}
\multicolumn{1}{c}{\multirow{-2}{*}{Model}} & \multirow{-2}{*}{Size} & Intro         & Inter         & Comp          & All           & Intro          & Inter         & Comp          & All           & Intro          & Inter          & Comp           & All            \\\hline
Codex                                       & 12B                    & 4.14          & 0.14          & 0.02          & 0.92          & 9.65           & 0.51          & 0.09          & 2.25          & -          & -           & -           & -           \\
GPT2                                        & 1.5B                   & 1.30          & 0.70          & 0.00          & 0.68          & 3.60           & 1.03          & 0.00          & 1.34          & -              & -              & -              & -              \\
GPT-Neo                                     & 2.7B                   & 3.90          & 0.57          & 0.00          & 1.12          & 5.50           & 0.80          & 0.00          & 1.58          & -              & -              & -              & -              \\
GPT-J                                       & 6B                     & 5.60          & 1.00          & 0.50          & 1.82          & 9.20           & 1.73          & 1.00          & 3.08          & -              & -              & -              & -              \\
StarCoder                                     & 164M        & 1.73  & 0.44  & 0.01  & 0.63  & 4.70  & 1.43  & 0.46  & 1.89  & 14.80  & 5.50  & 3.80  & 7.02  \\
CodeGen                                     & 350M        & 1.54  & 0.38  & 0.08  & 0.56  & 4.91  & 1.26  & 0.37  & 1.82  & 17.40  & 5.51  & 4.10  & 7.62  \\
CodeT5                                      & 220M                   & 0.71          & 0.27          & 0.03          & 0.31 
         & 2.40           & 0.94          & 0.12          & 1.07          & 9.90           & 3.53           & 1.60     
 & 4.42           \\ 
CodeT5+                                      & 770M                   & 4.41          & 0.99          & 0.26          & 1.53
         & 9.92           & 2.59          & 1.05          & 3.75          & 23.50           & 8.33           & 6.50     
 & 11.00           \\ \hline
CodeT5                                      & 770M                   & 3.30          & 0.68          & 0.15          & 1.10          & 8.12           & 1.89          & 0.74          & 2.91          & 21.30          & 6.53           & 6.00           & 9.38           \\ \rowcolor{gray!20}
CodeT5+CodePLAN                                 & 770M                   & \textbf{7.87} & \textbf{1.61} & \textbf{0.42} & \textbf{2.62} & \textbf{14.66} & \textbf{3.54} & \textbf{1.59} & \textbf{5.37} & \textbf{28.60} & \textbf{9.17} & \textbf{8.60} & \textbf{12.94} \\ \hline
\multicolumn{2}{c|}{Relative Improvement }       & \textcolor{red}{138.5\%}       & \textcolor{red}{136.8\%}       & \textcolor{red}{162.5\%}     & \textcolor{red}{138.2\%}    & \textcolor{red}{80.5\%}    & \textcolor{red}{87.3\%}    & \textcolor{red}{114.9\%}    & \textcolor{red}{84.5\%}   & \textcolor{red}{34.3\%}   & \textcolor{red}{40.4\%}    & \textcolor{red}{43.3\%}     & \textcolor{red}{38.0\%}  \\ \hline
\end{tabular}
}
\caption{Performance by Pass@k on APPS: ``Intro'': introductory, ``Inter'': interview, ``Comp'': competition.}
\label{tab:tabel1}
\end{table*}

%% file: tabels/03_table_3.tex
\begin{table*}[t]

\centering
\resizebox{0.9\textwidth}{!} {
\begin{tabular}{c|cccc|cccc}
\hline
\multicolumn{1}{c|}{}   & \multicolumn{4}{c|}{\emph{Pass@1}}     & \multicolumn{4}{c}{\emph{Pass@5}}   \\ 
\cline{2-9} \multicolumn{1}{c|}{\multirow{-2}{*}{Method}} & Intro         & Inter         & Comp          & All           & Intro          & Inter         & Comp          & All \\
\hline
\multicolumn{9}{c}{\cellcolor[HTML]{EFEFEF}CodeGen-350M} \\
\hline
\multicolumn{1}{c|}{standard finetune} & 1.54 & 0.38 & 0.08 & 0.56 & 4.91 & 1.26 & 0.37 & 1.82 \\
\hline
\multicolumn{1}{c|}{CoT finetune} & 1.38 & 0.35 & 0.06 & 0.50 & 3.95 & 1.21 & 0.21 & 1.56 \\
\hline
\multicolumn{1}{c|}{CodePLAN w/o PS} & \textbf{2.06}  & \textbf{0.56}  & \textbf{0.09}  & \textbf{0.77}  & \textbf{5.27}  & \textbf{1.62}  & \textbf{0.40}  & \textbf{2.11} \\
\hline
\multicolumn{9}{c}{\cellcolor[HTML]{EFEFEF}CodeT5-770M} \\
\hline
\multicolumn{1}{c|}{standard finetune} & 3.30 & 0.68 & 0.15 & 1.10 & 8.12 & 1.89 & 0.74 & 2.91 \\
\hline
\multicolumn{1}{c|}{CoT finetune} & 3.22 & 0.78 & 0.14 & 1.15 & 7.65 & 1.99 & 0.36 & 2.84 \\
\hline
\multicolumn{1}{c|}{CodeRL*} & 3.76 & 0.79 & 0.16 & 1.25 & 9.20 & 2.08 & 0.69 & 3.22 \\
\hline
\multicolumn{1}{c|}{CodePLAN w/o PS} & \textbf{3.90}  & \textbf{0.80}  & \textbf{0.20}  & \textbf{1.30}  & \textbf{9.25}  & \textbf{2.17}  & \textbf{0.78}  & \textbf{3.31} \\
\hline
\end{tabular}
}
\caption{Results with different training methods on APPS.} 
\label{tab: tabel3}
\end{table*}

%% file: tabels/04_table_4.tex
\begin{table}[t]

\centering
\resizebox{0.45\textwidth}{!} {
\begin{tabular}{c|c|c|c}
\hline
\multicolumn{1}{c|}{Method}   & \multicolumn{1}{c}{\emph{Pass@1}}     & \multicolumn{1}{c}{\emph{Pass@5}}    & \multicolumn{1}{c}{\emph{Pass@80}} \\ 
\hline
\multicolumn{4}{c}{\cellcolor[HTML]{EFEFEF}CodeGen-350M} \\
\hline
\multicolumn{1}{c|}{Standard finetune} & \multicolumn{1}{c}{7.51}  & \multicolumn{1}{c}{14.98}  & \multicolumn{1}{c}{30.29} \\
\hline
\multicolumn{1}{c|}{CoT finetune} & \multicolumn{1}{c}{7.95} & \multicolumn{1}{c}{15.01} & \multicolumn{1}{c}{30.12}   \\
\hline
\multicolumn{1}{c|}{CodePLAN w/o PS} & \multicolumn{1}{c}{\textbf{10.39}}  & \multicolumn{1}{c}{\textbf{18.66}}  & \multicolumn{1}{c}{\textbf{33.05}}   \\
\hline
\multicolumn{4}{c}{\cellcolor[HTML]{EFEFEF}CodeT5-770M} \\
\hline
\multicolumn{1}{c|}{Standard finetune} & \multicolumn{1}{c}{13.78}  & \multicolumn{1}{c}{26.01}  & \multicolumn{1}{c}{47.89} \\
\hline
\multicolumn{1}{c|}{CoT finetune} & \multicolumn{1}{c}{12.06} & \multicolumn{1}{c}{24.03} & \multicolumn{1}{c}{47.01}   \\
\hline
\multicolumn{1}{c|}{CodePLAN w/o PS} & \multicolumn{1}{c}{\textbf{15.13}}  & \multicolumn{1}{c}{\textbf{28.07}}  & \multicolumn{1}{c}{\textbf{51.09}}   \\
\hline
\end{tabular}
}
\caption{Results with different training methods on MBPP.} 
\label{tab: Tabel 4}
\end{table}

%% file: tabels/05_table_5.tex
\begin{table*}[t]
\centering
\small

\resizebox{1.0\textwidth}{!} {
\begin{tabular}{c|cccc|cccc|cccc}
\hline
\multicolumn{1}{c|}{\multirow{-0.5}{*}{Plan Sampling Number}}   & \multicolumn{4}{c|}{\emph{Pass@1}}  & \multicolumn{4}{c|}{\emph{Pass@5}}  & \multicolumn{4}{c}{\emph{Pass@100}}   \\ 
\cline{2-13}
\multicolumn{1}{c|}{} & Intro         & Inter         & Comp          & All           & Intro          & Inter         & Comp          & All           & Intro          & Inter          & Comp           & All            \\\hline
\multicolumn{1}{c|}{N=0}  & 3.82  & 0.78  & 0.15  & 1.26  & 9.25  & 2.07 & 0.68 & 3.22 & 21.30 & 6.53 & 6.00 & 9.38 \\ 
\multicolumn{1}{c|}{N=1}  & 2.40  & 0.57  & 0.09  & 0.84  & 5.49  & 1.50 & 0.43 & 2.08 & 15.10 & 5.43 & 3.90 & 7.06 \\ 
\multicolumn{1}{c|}{N=5}  & 5.33  & 1.04  & 0.24  & 1.74  & 11.02  & 2.59 & 1.04 & 3.96 & 24.70 & 7.57 & 7.10 & 10.90 \\ 
\multicolumn{1}{c|}{N=10}  & 6.61  & 1.33  & 0.35  & 2.19  & 12.92  & 3.03 & 1.33 & 4.67 & 26.30 & 8.43 & 8.10 & 11.94 \\ 
\rowcolor{gray!20}
\multicolumn{1}{c|}{N=20}  & \textbf{7.87}  & \textbf{1.61}  & \textbf{0.42}  & \textbf{2.62}  & \textbf{14.66}  & \textbf{3.54} & \textbf{1.59} & \textbf{5.37} & \textbf{28.60}  & \textbf{9.17} & \textbf{8.60} & \textbf{12.94} \\ \hline
\end{tabular}
}
\caption{Results of ablation experiments with different number of sampling plans.}
\label{tab: Tabel 5}
\end{table*}

%% file: tabels/07_table_7.tex
\begin{table}[t]
\vspace{-1mm}

\centering
\resizebox{0.45\textwidth}{!} {
\begin{tabular}{c|c|c|c}
\hline
\multicolumn{1}{c|}{Method}   & \multicolumn{1}{c}{\emph{Pass@1}}     & \multicolumn{1}{c}{\emph{Pass@5}}    & \multicolumn{1}{c}{\emph{Pass@10}} \\ 
\hline
\multicolumn{4}{c}{\cellcolor[HTML]{EFEFEF}CodeT5-770M} \\
\hline
\multicolumn{1}{c|}{Without Plan} & \multicolumn{1}{c}{0.37}  & \multicolumn{1}{c}{1.19}  & \multicolumn{1}{c}{1.71} \\
\hline
\multicolumn{1}{c|}{Problem to Plan} & \multicolumn{1}{c}{0.73} & \multicolumn{1}{c}{2.02} & \multicolumn{1}{c}{2.81}   \\
\hline
\multicolumn{1}{c|}{Code to Plan} & \multicolumn{1}{c}{\textbf{1.35}}  & \multicolumn{1}{c}{\textbf{3.52}}  & \multicolumn{1}{c}{\textbf{4.78}}   \\
\hline
\end{tabular}
}
\vspace{-1mm}
\caption{Quality results of solution plans generated from LLM using different approaches on APPS.} 
\label{tab: Tabel 7}
\end{table}

%% file: sections/05_discussion.tex
\section{Discussion}
\paragraph{How Does Solution Plan Quality Impact Model Performance in Code Generation?}
Based on the data in Table~\ref{tab: Tabel 5} and Figure~\ref{fig:Number plans}, it is clear that the quality of solution plans significantly influences model performance. Instead of enhancing the model's code generation capabilities, subpar solution plans might actually degrade its performance. We believe that these lower-quality plans could be misconstrued by the model as noise, negatively affecting its foundational capabilities. On the other hand, high-quality solution plans can greatly boost the model's code generation, leading to a higher output of accurate codes. As such, devising a method to select high-quality solution plans becomes crucial in code generation, 
\vspace{-2mm}
\paragraph{What Distinguishes Our Approach from Conventional Code Post-Processing Methods in Code Generation?}

\begin{figure}
    \centering
    \includegraphics[width=0.85\linewidth]{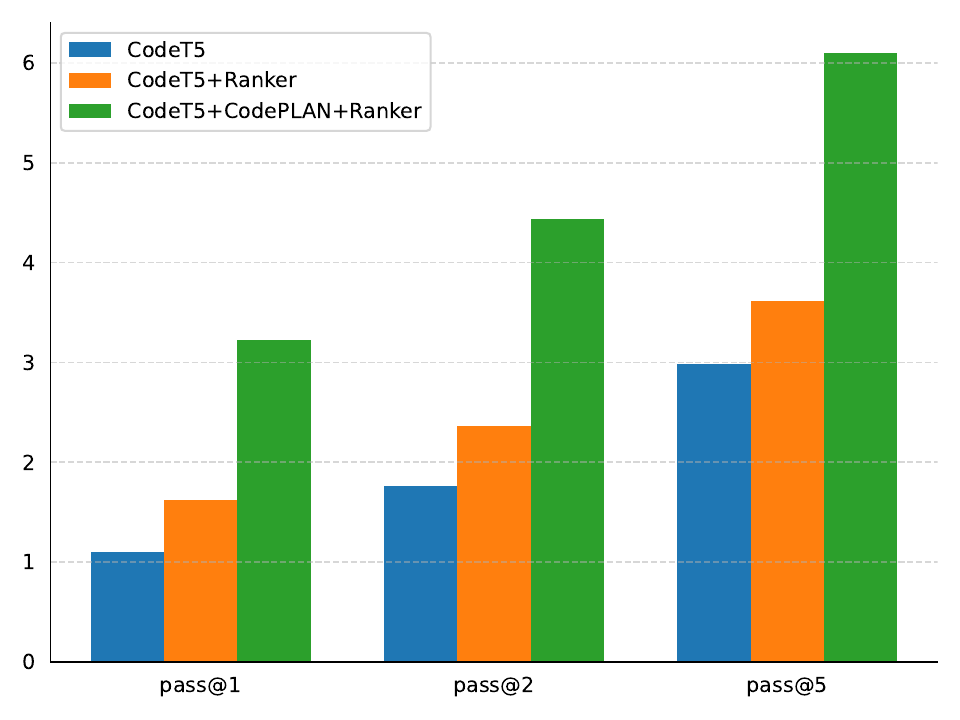}
    \vspace{-3mm}
    \caption{The complementarity between CodeRanker and our CodePLAN.}
    \label{fig: figure6}
    \vspace{-4mm}
\end{figure}

It's worth noting that various post-processing code methodologies~\citep{chen2022codet,zhang2022coder, inala2022fault} employ a technique to rank potential codes. However, such a ranking strategy doesn't inherently enhance the model's code generation capabilities. In contrast, our approach actively encourages the model to produce more correct codes. Consider a scenario in a programming competition: a conventionally fine-tuned model might generate 100 code samples for a problem, yet only 1 or 2 of those might pass the unit test. This low accuracy complicates the task of code ranking. Conversely, our method drives the model to yield a much higher proportion of accurate codes—perhaps 50 to 90 out of 100. This surge in accurate code generation certainly aids in the ranking process. We adopted the CodeRanker~\citep{inala2022fault} to train a Ranker to validate our assertions. Figure \ref{fig: figure6} presents the results of CodeT5, CodeT5+Ranker, and CodeT5+CodePLAN+Ranker on APPS. CodeT5+Ranker improves CodeT5's pass@\{1,2,5\} by an average of 54.5\%, and in combination with CodePLAN, CodeT5+Ranker+CodePLAN brings an average of 91.3\% improvement, it demonstrates the advantages of CodePLAN in code ranking tasks and it is also proved that the two are orthogonal.

%% file: sections/06_conclusion.tex
\section{Conclusion}
We introduced an innovative code generation framework named CodePLAN. During its training phase, CodePLAN uniquely emphasizes the generation of solution plans, aiming to refine and optimize the overall code generation process. In the inference phase, the framework leverages autonomously produced solution plans, strategically enhancing the likelihood of producing accurate codes. Our extensive experimental evaluations provide compelling evidence of the efficacy of our approach, demonstrating a significant boost in the performance of smaller models in code generation.

%% file: sections/07_limitations.tex
\section*{Limitations}
\label{sec: limitations}
Here we summarize two main limitations:

Firstly, the first limitation is that due to the limited dataset we have not considered what CodePLAN's preferences are for different types of programming topics. Some code generation datasets are now starting to consider dividing the dataset based on different algorithms~\citep{li2023taco}, so we may in the future integrate different algorithms in the solution plan to enhance CodePLAN's ability to learn different algorithms and evaluate CodePLAN at a fine-grained level (e.g., different algorithm types).

Secondly, we considered only one programming language. In our future work, we plan to explore the adaptability of this framework across different programming languages and more intricate coding scenarios. With the continuous advancement in automatic code generation techniques, we believe methods like CodePLAN will play a pivotal role in furthering the progress of this domain.

%% file: sections/08_acknowledgments.tex
\section*{Acknowledgments}
\label{sec: acknowledgments}

The work is supported in part by the Natural Science Foundation of Shandong Province, China (Grant No. ZR2021MF059), the National Natural Science Foundation of China (Grant Nos. 62192731, 62072007, 62192733, 61832009, 62192730), the National Key R\&D Program (Grant No. 2023YFB4503801) and the Key Program of Hubei (Grant No. JD2023008).

%% file: sections/09_appendice.tex
\section{Early Exploration Experiments}
\label{Appendice:A}
In the course of our preliminary experimental investigations, we discerned that the solution plans, conjured by LLM(specifically refers to gpt-3.5-turbo) from the problem descriptions, were integrated with these descriptions to serve as prompts. This composite data was subsequently fed to the fine-tuned smaller models. We observed that this methodology marginally enhanced the smaller models' aptitude in addressing complex programming problems. Furthermore, by manually selecting the most superior solution plans, we were able to significantly amplify the smaller models' code generation capacities. We singled out these high-quality solution plans based on the substantial improvement they afforded to the CodeT5-770M model. Interestingly, as revealed from our analysis of Table~\ref{tab: tabel_8}, these high-quality solution plans demonstrated a degree of generalization, thereby leading to a substantial improvement in the performance of other smaller models. This crucial insight spurred us to explore a fresh approach aimed at tapping into the latent potential of smaller models in tackling intricate programming problems.
\input{tabels/08_table_8}
\section{Examples}
\subsection{Examples of Generated Plans}
\label{sec: example plan}
\begin{figure*}
    \centering
    \includegraphics[width=0.98\linewidth]{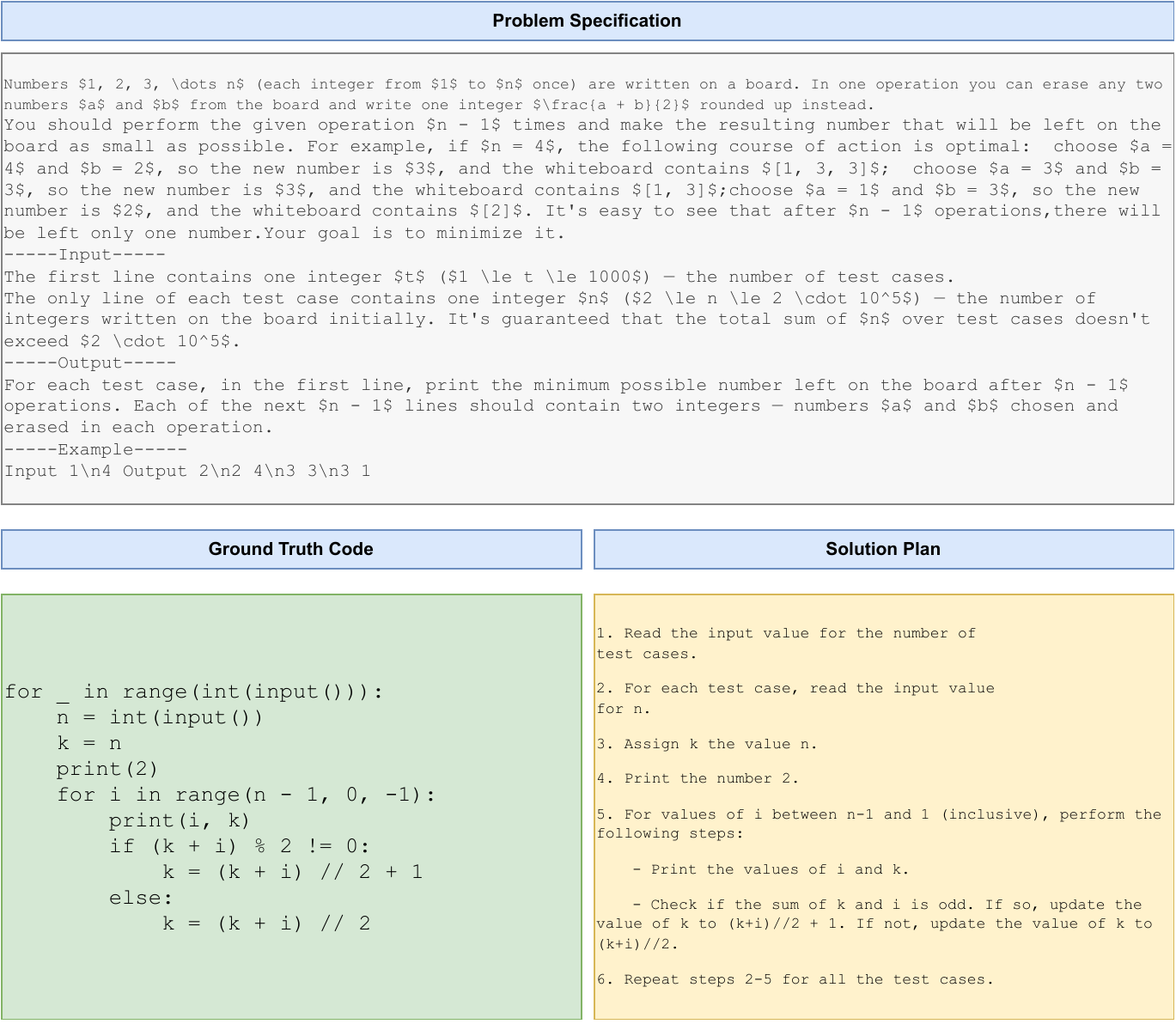}
    \vspace{-3mm}
    \caption{An example of an interview-level problem and ground truth code on the APPS benchmark, and a solution plan generated by LLM based on the ground truth code.}
    \label{fig:figure_7}
    \vspace{-4mm}
\end{figure*}

\begin{figure*}
    \centering
    \includegraphics[width=0.98\linewidth]{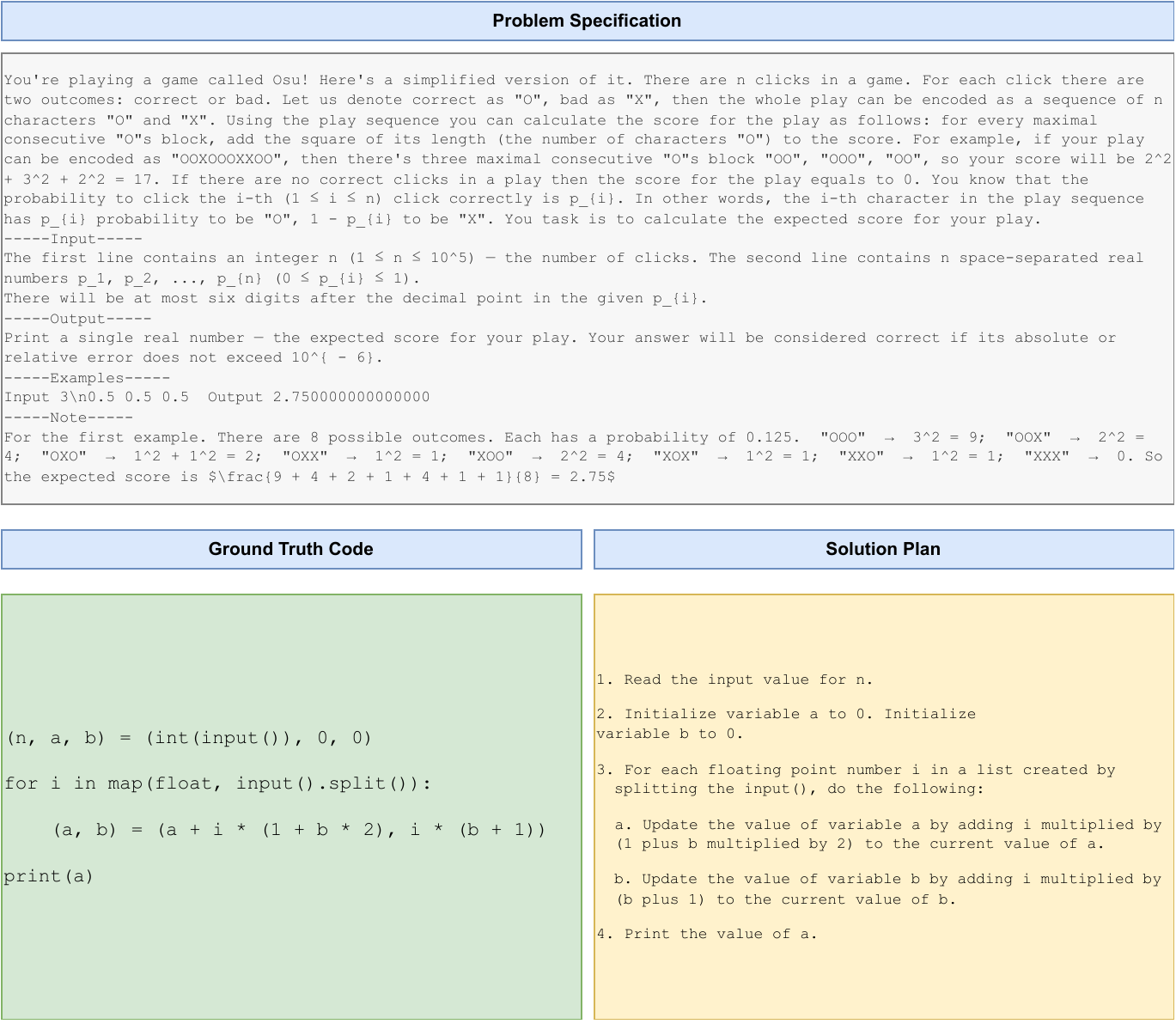}
    \vspace{-3mm}
    \caption{An example of an competition-level problem and ground truth code on the APPS benchmark, and a solution plan generated by LLM based on the ground truth code.}
    \label{fig:figure_8}
    \vspace{-4mm}
\end{figure*}

Some examples of solution plans generated by LLM based on ground truth codes are shown in Figure~\ref{fig:figure_7} and Figure~\ref{fig:figure_8}. The example in Figure ~\ref{fig:figure_7} is an interview-level problem on the APPS benchmark, and the example in Figure~\ref{fig:figure_8} is a competition-level problem on the APPS benchmark.
\subsection{Example Generated Programs}
\label{sec: example programs}
\begin{figure*}
    \centering
    \includegraphics[width=0.98\linewidth]{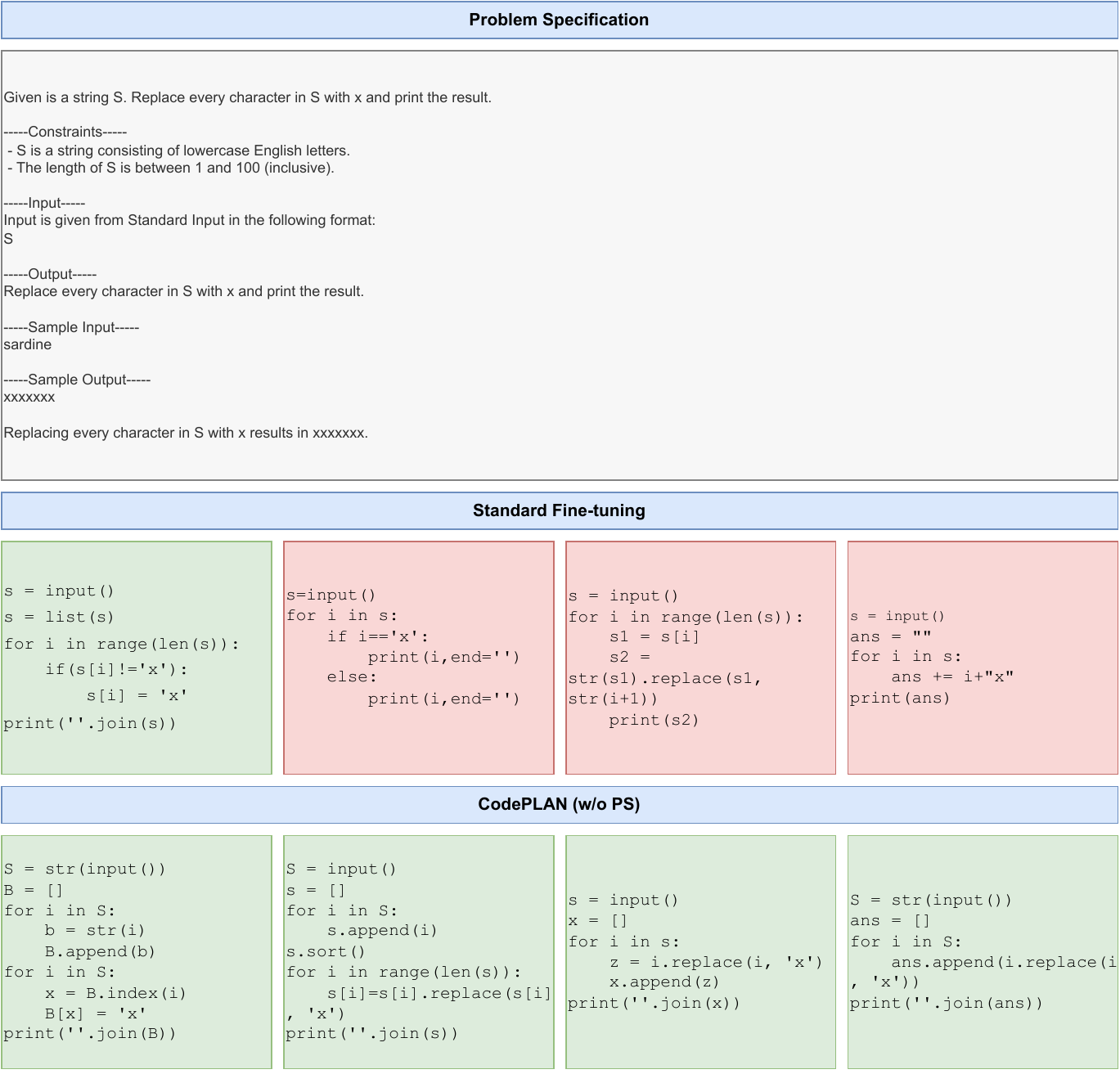}
    \vspace{-3mm}
    \caption{\textbf{An example of a problem on the APPS benchmark and code generated by CodeT5 with different ways of fine-tuning:} only one of the four copies of code generated by CodeT5 with standard finetune passes the unit test, while all four copies of code generated by CodeT5 with our CodePLAN (w/o PS) fine-tuning pass the unit test.}
    \label{fig:figure_9}
    \vspace{-4mm}
\end{figure*}

\begin{figure*}
    \centering
    \includegraphics[width=0.98\linewidth]{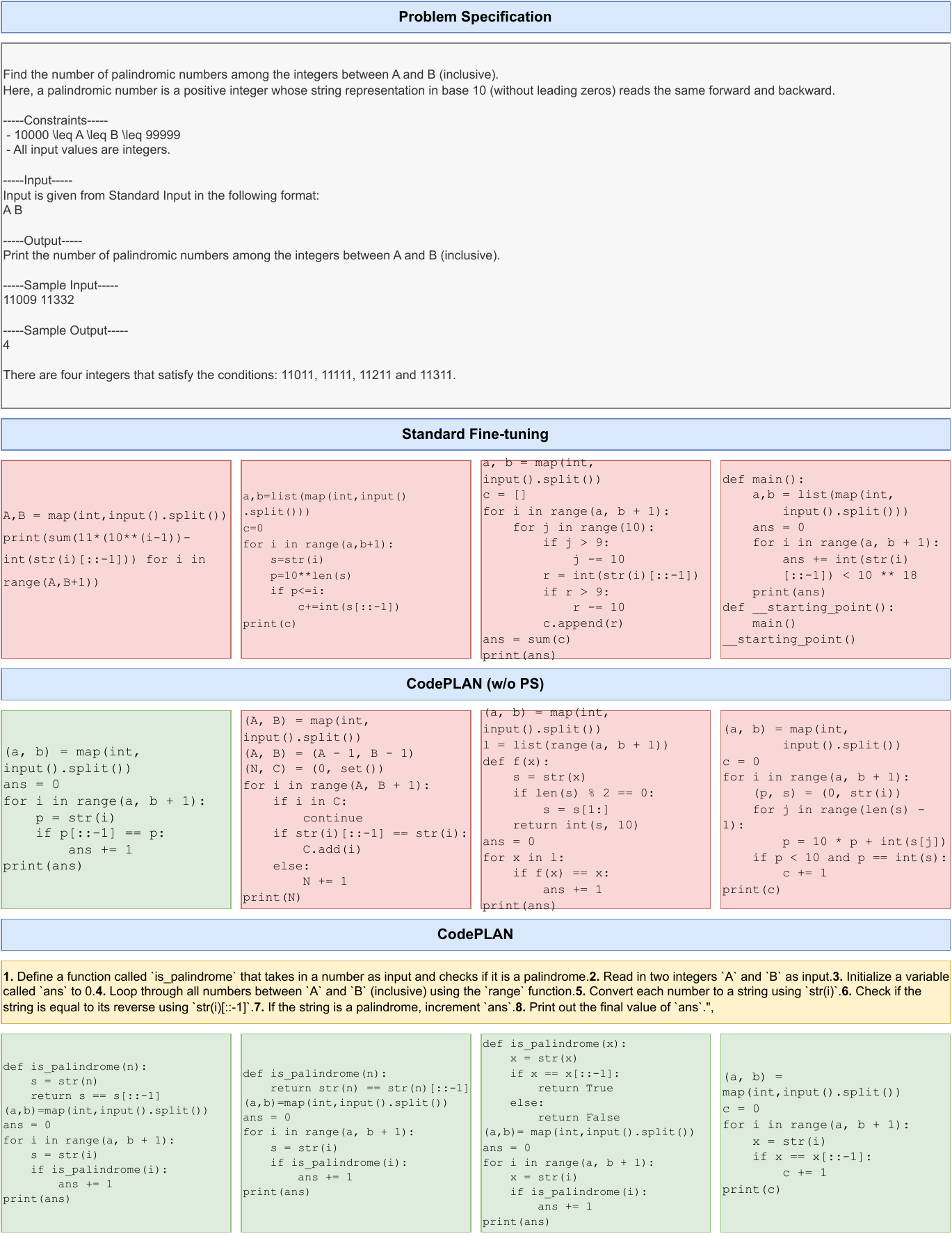}
    \vspace{-3mm}
    \caption{\textbf{An example of a problem on the APPS benchmark and code generated by CodeT5 with different ways of fine-tuning:} CodeT5 with the standard fine-tune method generates four copies of code that do not pass the unit test, while CodeT5 with our CodePLAN (w/o PS) generates one of the four copies of code that pass the unit test, and we select a high-quality solution plan to be spliced after the problem description as a prompt, and CodeT5 with CodePLAN generates four copies of code that pass the unit test.}
    \label{fig:figure_10}
    \vspace{-4mm}
\end{figure*}

\begin{figure*}
    \centering
    \includegraphics[width=0.98\linewidth]{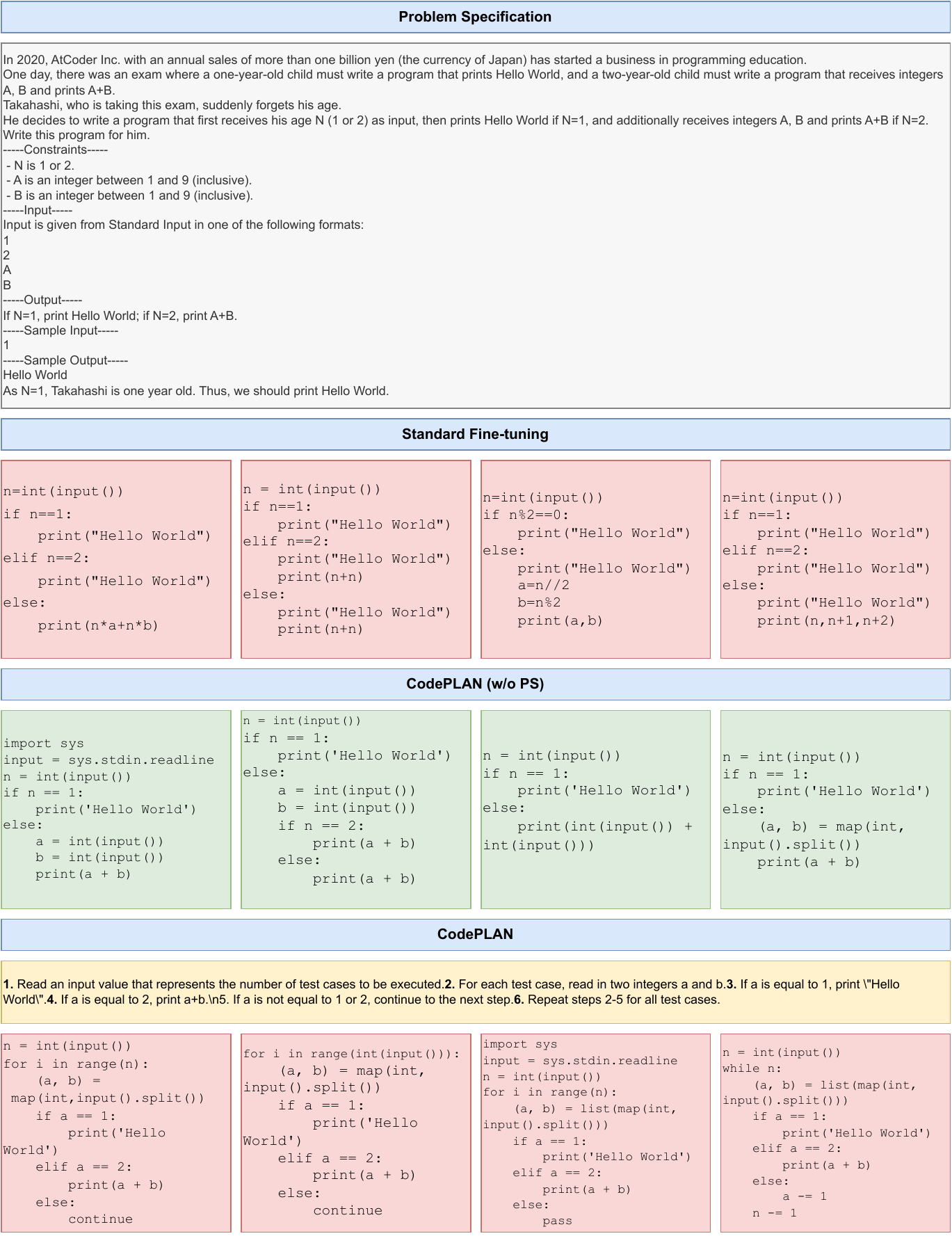}
    \vspace{-3mm}
    \caption{\textbf{An example of a problem on the APPS benchmark and code generated by CodeT5 with different ways of fine-tuning:} All four copies of CodeT5 generated by the standard fine-tuning method fail the unit test, while all four copies of CodeT5 generated by our CodePLAN (w/o PS) pass the unit test. However, we select an incorrect solution plan and splice it after the problem description as a prompt that all four copies of CodeT5 with CodePLAN do not pass the unit test.}
    \label{fig:figure_11}
    \vspace{-4mm}
\end{figure*}

Figures~\ref{fig:figure_9} to~\ref{fig:figure_11} provide illustrative examples of code produced by CodeT5, under the influence of various fine-tuning methods. Code segments failing to pass the unit test are presented on a red background, whilst code that successfully navigates the unit test is outlined on a green background. As evidenced in Figure~\ref{fig:figure_9}, CodeT5, when fine-tuned using our proposed approach, generates a significantly higher volume of correct code than when it is fine-tuned with the standard method. Figure~\ref{fig:figure_10} depicts how CodeT5, when fine-tuned using our methodology, can effectively address issues that remain unresolved when CodeT5 is fine-tuned using the standard method. This outcome stems from the standard fine-tuning method's limited success in enhancing the smaller model's reasoning abilities, whereas our method successfully amplifies the mini-model's cognitive capabilities by fostering mutual enhancement between the two tasks. Furthermore, our strategy of deploying high-quality solution plans as cues significantly augments the probability of correct code generation. Figure ~\ref{fig:figure_11}, however, reveals a failure scenario where despite our method yielding a greater number of accurate codes, the selection of an incorrect solution plan (generated by the greedy decoding of CodePLAN w/o PS) as a cue fails to produce correct codes. This is attributable to the incorrect solution plan being perceived as noise by the smaller model, thereby compromising the smaller model's innate abilities.

%% file: tabels/08_table_8.tex
\begin{table*}[t]

\centering
\resizebox{0.95\textwidth}{!} {
\begin{tabular}{c|cccc|}
\hline
\multicolumn{1}{c|}{Method}   & \multicolumn{1}{c}{Intro}         & \multicolumn{1}{c}{Inter}         & \multicolumn{1}{c}{Comp}          & \multicolumn{1}{c}{All}   \\ 
\hline
\multicolumn{5}{c}{\cellcolor[HTML]{EFEFEF}\textbf{CodeT5-770M\quad \quad \quad \quad Pass@1}} \\
\hline
\multicolumn{1}{c|}{No-Plan} & 3.30 & 0.68 & 0.16 & \multicolumn{1}{c}{1.10}  \\
\multicolumn{1}{c|}{With-Plan(greedy)} & $4.90 \ \ (\textcolor{red}{\uparrow 48.5\%})$ & $1.03 \ \ (\textcolor{red}{\uparrow 51.5\%})$ & $0.14 \ \ (\textcolor{blue}{\downarrow 14.3\%})$ & \multicolumn{1}{c}{$1.62 \ \ (\textcolor{red}{\uparrow 47.3\%})$} \\
\multicolumn{1}{c|}{With-Plan(best)} & $\textbf{9.72} \ \ (\textcolor{red}{\uparrow 194.5\%})$  & $\textbf{1.91} \ \ (\textcolor{red}{\uparrow 180.9\%})$  & $\textbf{0.45} \ \ (\textcolor{red}{\uparrow 181.3\%})$  & \multicolumn{1}{c}{$\textbf{3.18} \ \ (\textcolor{red}{\uparrow 189.1\%})$} \\
\hline
\multicolumn{5}{c}{\cellcolor[HTML]{EFEFEF}\textbf{CodeT5-770M\quad \quad \quad \quad Pass@5}} \\
\hline
\multicolumn{1}{c|}{No-Plan} & 8.12 & 2.01 & 0.74 & \multicolumn{1}{c}{2.98}  \\
\multicolumn{1}{c|}{With-Plan(greedy)} & $11.30 \ \ (\textcolor{red}{\uparrow 39.2\%})$ & $2.69 \ \ (\textcolor{red}{\uparrow 33.8\%})$ & $0.61 \ \ (\textcolor{blue}{\downarrow 17.6\%})$ & \multicolumn{1}{c}{$4.00 \ \ (\textcolor{red}{\uparrow 34.2\%})$} \\
\multicolumn{1}{c|}{With-Plan(best)} & $\textbf{18.90} \ \ (\textcolor{red}{\uparrow 132.8\%})$  & $\textbf{4.56} \ \ (\textcolor{red}{\uparrow 126.9\%})$  & $\textbf{1.80} \ \ (\textcolor{red}{\uparrow 143.2\%})$  & \multicolumn{1}{c}{$\textbf{6.88} \ \ (\textcolor{red}{\uparrow 130.9\%})$} \\
\hline
\multicolumn{5}{c}{\cellcolor[HTML]{EFEFEF}\textbf{CodeT5-220M\quad \quad \quad \quad Pass@1}} \\
\hline
\multicolumn{1}{c|}{No-Plan} & 0.71 & 0.27 & 0.03 & \multicolumn{1}{c}{0.31}  \\
\multicolumn{1}{c|}{With-Plan(greedy)} & $1.36 \ \ (\textcolor{red}{\uparrow 91.5\%})$ & $ 0.29 \ \ (\textcolor{red}{\uparrow 7.41\%})$ & $0.04 \ \ (\textcolor{red}{\uparrow 33.3\%})$ & \multicolumn{1}{c}{$0.45 \ \ (\textcolor{red}{\uparrow 45.2\%})$} \\
\multicolumn{1}{c|}{With-Plan(best)} & $\textbf{1.62} \ \ (\textcolor{red}{\uparrow 128.2\%})$  & $\textbf{0.40} \ \ (\textcolor{red}{\uparrow 48.2\%})$  & $\textbf{0.06} \ \ (\textcolor{red}{\uparrow 100.0\%})$  & \multicolumn{1}{c}{$\textbf{0.58} \ \ (\textcolor{red}{\uparrow 87.1\%})$} \\
\hline
\multicolumn{5}{c}{\cellcolor[HTML]{EFEFEF}\textbf{CodeT5-220M\quad \quad \quad \quad Pass@5}} \\
\hline
\multicolumn{1}{c|}{No-Plan} & 2.40 & 0.94 & 0.12 & \multicolumn{1}{c}{1.07}  \\
\multicolumn{1}{c|}{With-Plan(greedy)} & $3.56 \ \ (\textcolor{red}{\uparrow 48.3\%})$ & $1.02 \ \ (\textcolor{red}{\uparrow 8.5\%})$ & $0.19 \ \ (\textcolor{red}{\uparrow 58.3\%})$ & \multicolumn{1}{c}{$1.36 \ \ (\textcolor{red}{\uparrow 27.1\%})$} \\
\multicolumn{1}{c|}{With-Plan(best)} & $\textbf{4.71} \ \ (\textcolor{red}{\uparrow 96.3\%})$  & $\textbf{1.22} \ \ (\textcolor{red}{\uparrow 29.8\%})$  & $\textbf{0.24} \ \ (\textcolor{red}{\uparrow 100.0\%})$  & \multicolumn{1}{c}{$\textbf{1.70} \ \ (\textcolor{red}{\uparrow 58.9\%})$} \\
\hline
\multicolumn{5}{c}{\cellcolor[HTML]{EFEFEF}\textbf{GPT2-1.5B\quad \quad \quad \quad \quad \quad Pass@1}} \\
\hline
\multicolumn{1}{c|}{No-Plan} & 1.30 & \textbf{0.70} & 0.00 & \multicolumn{1}{c}{0.68}  \\
\multicolumn{1}{c|}{With-Plan(greedy)} & $3.37 \ \ (\textcolor{red}{\uparrow 159.2\%})$ & $0.50 \ \ (\textcolor{blue}{\downarrow 40.0\%})$ & $0.04 \ \ (\textcolor{red}{\uparrow \infty})$ & \multicolumn{1}{c}{$0.94 \ \ (\textcolor{red}{\uparrow 38.2\%})$} \\
\multicolumn{1}{c|}{With-Plan(best)} & $\textbf{4.13} \ \ (\textcolor{red}{\uparrow 217.7\%})$  & $0.65 \ \ (\textcolor{blue}{\downarrow 7.1\%})$  & $\textbf{0.08} \ \ (\textcolor{red}{\uparrow \infty})$  & \multicolumn{1}{c}{$\textbf{1.20} \ \ (\textcolor{red}{\uparrow 76.5\%})$} \\
\hline
\multicolumn{5}{c}{\cellcolor[HTML]{EFEFEF}\textbf{GPT2-1.5B\quad \quad \quad \quad \quad \quad Pass@5}} \\
\hline
\multicolumn{1}{c|}{No-Plan} & 3.60 & 1.03 & 0.00 & \multicolumn{1}{c}{1.34}  \\
\multicolumn{1}{c|}{With-Plan(greedy)} & $8.20 \ \ (\textcolor{red}{\uparrow 127.8\%})$ & $1.54 \ \ (\textcolor{red}{\uparrow 49.5\%})$ & $0.22 \ \ (\textcolor{red}{\uparrow \infty})$ & \multicolumn{1}{c}{$2.51 \ \ (\textcolor{red}{\uparrow 87.3\%})$} \\
\multicolumn{1}{c|}{With-Plan(best)} & $\textbf{9.66} \ \ (\textcolor{red}{\uparrow 168.3\%})$  & $\textbf{1.79} \ \ (\textcolor{red}{\uparrow 73.8\%})$  & $\textbf{0.38} \ \ (\textcolor{red}{\uparrow \infty})$  & \multicolumn{1}{c}{$\textbf{3.02} \ \ (\textcolor{red}{\uparrow 125.4\%})$} \\
\hline
\end{tabular}
}
\caption{Results of different smaller models on APPS with different solution plans generated from LLM.} 
\label{tab: tabel_8}
\end{table*}